\documentclass[12pt]{article}
\usepackage{html}
\usepackage{epsf}
\usepackage[pdftex]{graphicx}
\usepackage{graphicx}
\usepackage{amssymb}
\usepackage{amsmath}
\usepackage{hyperref}
\begin{document}
\begin{titlepage}
\includegraphics[width=150mm]{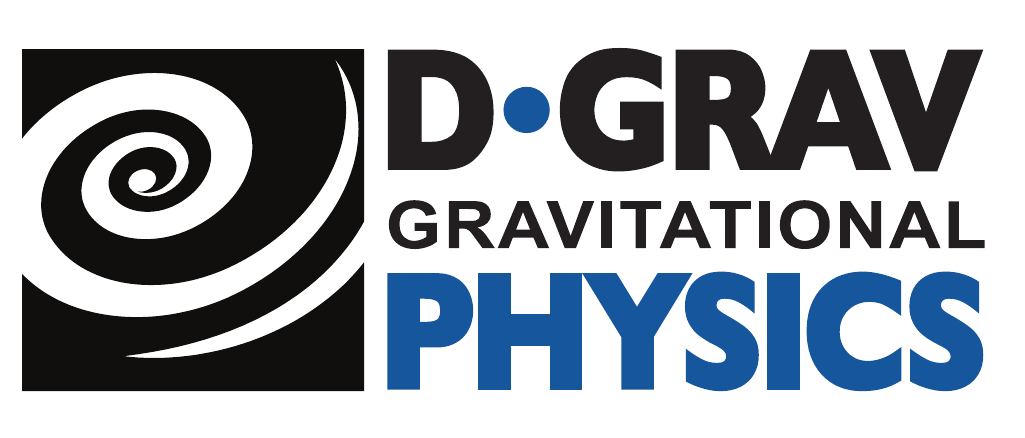}
\begin{center}
{ \Large {\bf MATTERS OF GRAVITY}}\\ 
\bigskip
\hrule
\medskip
{The newsletter of the Division of Gravitational Physics of the American Physical 
Society}\\
\medskip
{\bf Number 52 \hfill December 2018}
\end{center}
\begin{flushleft}
\tableofcontents
\end{flushleft}
\end{titlepage}
\vfill\eject
\begin{flushleft}
\section*{\noindent  Editor\hfill}
David Garfinkle\\
\smallskip
Department of Physics
Oakland University
Rochester, MI 48309\\
Phone: (248) 370-3411\\
Internet: 
\htmladdnormallink{\protect {\tt{garfinkl-at-oakland.edu}}}
{mailto:garfinkl@oakland.edu}\\
WWW: \htmladdnormallink
{\protect {\tt{http://www.oakland.edu/physics/Faculty/david-garfinkle}}}
{http://www.oakland.edu/physics/Faculty/david-garfinkle}\\

\section*{\noindent  Associate Editor\hfill}
Greg Comer\\
\smallskip
Department of Physics and Center for Fluids at All Scales,\\
St. Louis University,
St. Louis, MO 63103\\
Phone: (314) 977-8432\\
Internet:
\htmladdnormallink{\protect {\tt{comergl-at-slu.edu}}}
{mailto:comergl@slu.edu}\\
WWW: \htmladdnormallink{\protect {\tt{http://www.slu.edu/arts-and-sciences/physics/faculty/comer-greg.php}}}
{http://www.slu.edu/arts-and-sciences/physics/faculty/comer-greg.php}\\
\bigskip
\hfill ISSN: 1527-3431


\bigskip

DISCLAIMER: The opinions expressed in the articles of this newsletter represent
the views of the authors and are not necessarily the views of APS.
The articles in this newsletter are not peer reviewed.

\begin{rawhtml}
<P>
<BR><HR><P>
\end{rawhtml}
\end{flushleft}
\pagebreak
\section*{Editorial}

The next newsletter is due June 2019.  Issues {\bf 28-52} are available on the web at
\htmladdnormallink 
{\protect {\tt {https://files.oakland.edu/users/garfinkl/web/mog/}}}
{https://files.oakland.edu/users/garfinkl/web/mog/} 
All issues before number {\bf 28} are available at
\htmladdnormallink {\protect {\tt {http://www.phys.lsu.edu/mog}}}
{http://www.phys.lsu.edu/mog}

Any ideas for topics
that should be covered by the newsletter should be emailed to me, or 
Greg Comer, or
the relevant correspondent.  Any comments/questions/complaints
about the newsletter should be emailed to me.

A hardcopy of the newsletter is distributed free of charge to the
members of the APS Division of Gravitational Physics upon request (the
default distribution form is via the web) to the secretary of the
Division.  It is considered a lack of etiquette to ask me to mail
you hard copies of the newsletter unless you have exhausted all your
resources to get your copy otherwise.

\hfill David Garfinkle 

\bigbreak

\vspace{-0.8cm}
\parskip=0pt
\section*{Correspondents of Matters of Gravity}
\begin{itemize}
\setlength{\itemsep}{-5pt}
\setlength{\parsep}{0pt}
\item Daniel Holz: Relativistic Astrophysics,
\item Bei-Lok Hu: Quantum Cosmology and Related Topics
\item Veronika Hubeny: String Theory
\item Pedro Marronetti: News from NSF
\item Luis Lehner: Numerical Relativity
\item Jim Isenberg: Mathematical Relativity
\item Katherine Freese: Cosmology
\item Lee Smolin: Quantum Gravity
\item Cliff Will: Confrontation of Theory with Experiment
\item Peter Bender: Space Experiments
\item Jens Gundlach: Laboratory Experiments
\item Warren Johnson: Resonant Mass Gravitational Wave Detectors
\item David Shoemaker: LIGO 
\item Stan Whitcomb: Gravitational Wave detection
\item Peter Saulson and Jorge Pullin: former editors, correspondents at large.
\end{itemize}
\section*{Division of Gravitational Physics (DGRAV) Authorities}
Chair: Emanuele Berti; Chair-Elect: 
Gary Horowitz; Vice-Chair: Nicolas Yunes. 
Secretary-Treasurer: Geoffrey Lovelace; Past Chair:  Peter Shawhan; Councilor: Beverly Berger
Members-at-large:
Kelly Holley-Bockelmann, Leo Stein, Lisa Barsotti, Theodore Jacobson, Michael Lam, Jess McIver.
Student Members: Cody Messick, Belinda Cheeseboro.
\parskip=10pt

\vfill\eject

\section*{\centerline
{we hear that \dots}}
\addtocontents{toc}{\protect\medskip}
\addtocontents{toc}{\bf DGRAV News:}
\addcontentsline{toc}{subsubsection}{
\it we hear that \dots , by David Garfinkle}
\parskip=3pt
\begin{center}
David Garfinkle, Oakland University
\htmladdnormallink{garfinkl-at-oakland.edu}
{mailto:garfinkl@oakland.edu}
\end{center}

Jocelyn Bell Burnell has been awarded the Breakthrough Prize in Fundamental Physics

Abhay Ashtekar has been awarded the APS Einstein Prize.

Stanley Whitcomb has been awarded the APS Richard A. Isaacson Award in Gravitational-Wave Science.

Katherine Freese has been awarded the APS Julius Edgar Lilienfeld Prize.

Carlos Lousto has been awarded the APS Edward A. Bouchet Award.

Rana Adhikari, Lisa Barsotti, Henriette Elvang, Stephen Fulling, Salman Habib, Lawrence Kidder, Derek Kimball, Mikhail Medvedev, Ingrid Stairs, Andrew Strominger, and Ignacio Taboada have been elected APS Fellows.

Hearty Congratulations!

\vfill\eject
\section*{\centerline
{April APS meeting}
}
\addtocontents{toc}{\protect\medskip}
\addcontentsline{toc}{subsubsection}{
\it APS April Meeting, by David Garfinkle}
\parskip=3pt
\begin{center}
David Garfinkle, Oakland University
\htmladdnormallink{garfinkl-at-oakland.edu}
{mailto:garfinkl@oakland.edu}
\end{center}

We have a very exciting DGRAV related program at the upcoming APS meeting April 13-16 in Denver, Colorado.  Our Chair-elect, Gary Horowitz, did an excellent job of putting together this program.

\vskip0.25truein
The DGRAV sponsored invited sessions are\\

{\bf Looking at Supermassive Black Holes} \\
(co-sponsored with DAP)\\
Shep Doeleman, {\it  Event Horizon Telescope}\\
Avery Broderick, {\it Event Horizon Telescope: Theoretical Background and Interpretation}\\
Stefan Gillessen, {\it Latest News from the Galactic Center}\\

{\bf Quantum Aspects of Gravitation} \\
Eva Silverstein, {\it dS/dS and T bar T}\\
Daniel Jafferis, {\it Traversable Wormholes}\\
Daniel Carney, {\it Tabletop Experiments for Quantum Gravity}\\

{\bf Einstein Prize Talk and Developments in Gravitational Theory} \\
Abhay Ashtekar, {\it Some Unforseen Aspects of Electromagnetic and Gravitational Waves}\\
Eanna Flanagan, {\it Gravitational Backreaction on Cosmic Strings}\\
Mihalis Dafermos, {\it The Interior Structure of Dynamic Vacuum Black Holes Without Symmetry}\\

{\bf Isaacson Award Talk and Ground Based Gravitational Wave Astronomy}\\
Stanley Whitcomb, {\it Making LIGO Possible: a Technical History}\\
Katerina Chatziioannou, {\it The Current State of Gravitational Wave Astronomy with LIGO and Virgo}\\
Evan Hall, {\it The Next Generation of Ground Based Gravitational Wave Detectors}\\

{\bf Space Based Gravitational Wave Astronomy} \\
Emanuele Berti, {\it Low Frequency Gravitational Waves from Massive Black Holes: Implications for Fundamental Physics and Astrophysics }\\
Maura McLaughlin, {\it Pulsar Timing Arrays: Observations, Analysis, and Insights into Galaxy Growth and Evolution}\\
Thomas Kupfer, {\it Galactic LISA Sources and their Potential for Multi-messenger and Multi-wavelength Studies}\\

{\bf Progress in Theoretical Gravitational Wave Astronomy} \\
Oliver Tattersall, {\it Signatures of Modified Gravity in Black Hole Quasi-Normal Modes}\\
Felix Julie, {\it Two-body Problem in Modified Gravities and EOB Theory}\\
Tanja Hinderer, {\it Progress in Theoretical Gravitational-Wave Astronomy of Neutron Star Binaries}\\

{\bf Tests of General Relativity} \\
Frank Eisenhauer, {\it General Relativistic Effects in Stellar Orbits around the Galactic Center Black Hole}\\
Kent Yagi, {\it Gravitational Wave Tests of General Relativity: Present and Future}\\
Scott Ransom, {\it Recent and Future Tests of GR using Pulsar Systems}\\

{\bf Edward Bouchet Award Talk and Progress in Numerical Simulations of Compact Binaries} \\
(co-sponsored with DCOMP)\\
Carlos Lousto, {\it Insight on Gravitational Waves and Astrophysics from Simulations of Binary Black Holes}\\
David Radice, {\it Numerical Relativity Simulations of Neutron Star Mergers}\\
Vasileios Paschalidis, {\it Lessons and Future Prospects from the Interplay of Multimessenger Astronomy and Computational Gravity}\\

{\bf NICER and its Implications} \\
(co-sponsored with DAP)\\
Deepto Chakrabarty, {\it An Overview of NICER Observations of Neutron Stars and Black Holes}\\
Tod Strohmayer, {\it NICER Probes of Neutron Stars and Dense Matter}\\
Hector Okada-Da Silva, {\it Probing Extreme Gravity with NICER}\\

{\bf Detection and Modeling of Binary Neutron Star Collisions} \\
(co-sponsored with DNP)\\
Samaya Nissanke, {\it LIGO Measurements of BNS Mergers}\\
Maria Drout, {\it Electromagnetic Follow-up to BNS Mergers}\\
Charles Horowitz, {\it Neutron Star Mergers: a Theoretical View}\\

{\bf Centennial of the Eddington Eclipse Expedition} \\
(co-sponsored withe FHP)\\
Daniel Kennefick, {\it No Shadow of Doubt: Eddington, Einstein, and the 1919 Eclipse}\\
Jeffrey Crelinsten, {\it Einstein's Jury: the Race to Test Relativity}\\

{\bf Precision Searches for New Long Range Forces, from Quantum Sensors to Spacecraft} \\ 
(co-sponsored with GPMFC)\\
Joel Berge, {\it MICROSCOPE Mission: First Constraints on the Violation of the Weak Equivalence Principle by a Light Scalar Dilaton}\\
Xing Rong, {\it Searching for an Exotic Spin-dependent Interaction with a Single Electron Spin Quantum Sensor}\\
Kent Irwin, {\it Fundamental Limits of Electromagnetic Axion and Hidden Photon Dark Matter Searches}\\

\vfill\eject

\end{document}